\documentclass[%
 reprint,
 amsmath,amssymb,
 aps,
prb,
]{revtex4-1}

\usepackage{textcomp}
\usepackage{graphicx}
\usepackage{dcolumn}
\usepackage{bm}
\usepackage{tabularx} 
\usepackage{lipsum}
\usepackage{multirow}
\usepackage{times}
\usepackage{subfig}
\usepackage{xcolor}

\def\Cu{$Cu$\hspace{0.09cm}}
\def\O{$O$\hspace{0.09cm}}
\def\Ta{$Ta$\hspace{0.09cm}}
\def\Ag{$Ag$\hspace{0.09cm}}
\def\tantala{$Ta_{2}O_{5}$\hspace{0.09cm}}

\begin{document}

\preprint{APS/123-QED}

\title{Structure and charge transport of amorphous \Cu-doped \tantala : An \textit{ab initio} study}

\author{Rajendra Thapa}
\email{rt887917@ohio.edu}
\affiliation{Department of Physics and Astronomy, \\
Nanoscale and Quantum Phenomena Institute (NQPI),\\
Ohio University, Athens, Ohio 45701, United States of America}%

\author{Bishal Bhattarai}
\email{bb6fq@mst.edu}
\affiliation{Department of Physics, \\
Missouri University of Science and Technology, Rolla, Missouri 65409, United States of America }

\author{M. N. Kozicki}%
\email{michael.kozicki@asu.edu}
\affiliation{Department of Electrical, Computer and Energy Engineering \\
Arizona State University, Tempe, AZ 85287, United States of America}%

\author{Kashi N. Subedi }%
\email{ks173214@ohio.edu}
\affiliation{Department of Physics and Astronomy, \\
Ohio University, Athens, Ohio 45701, United States of America}%

\author{D. A. Drabold}%
\email{drabold@ohio.edu}
\affiliation{Department of Physics and Astronomy, \\
Ohio University, Athens, Ohio 45701, United States of America}%

\date{\today}

\begin{abstract}
In this paper, we present \textit{ab initio} computer models of \Cu-doped amorphous \tantala , a promising candidate for Conducting Bridge Random Access Memory (CBRAM) memory devices, and study the structural, electronic, charge transport and vibrational properties based on plane-wave density functional methods. We offer an atomistic picture of the process of phase segregation/separation between \Cu and \tantala subnetworks. Electronic calculations show that the models are conducting with extended Kohn-Sham orbitals around the Fermi level. In addition to that, we also characterize the electronic transport using the Kubo-Greenwood formula modified suitably to calculate the space-projected conductivity (SPC)~\cite{Kiran_PSS}. Our SPC calculations show that \Cu clusters and under-coordinated \Ta adjoining the \Cu  are the conduction-active parts of the network. We also report information about the dependence of the electrical conductivity on the connectivity of the \Cu submatrix. Vibrational calculations for one of the models has been undertaken with an emphasis on localization and animation of representative modes. 
\end{abstract}

\maketitle

\section{\label{sec:level1}INTRODUCTION }

Novel non-volatile memory devices are an area of active inquiry. Research on ferroelectric random access memory (FeRAM)
and magnetoresistive random access memory (MRAM) have been limited due to underlying technological and scalability problems~\cite{Valov_2011}. Meanwhile, study of 
non-volatile memory, based on electrically switchable resistance or resistive random access memory (ReRAM), has gained considerable interest. The first studies of such 
resistive switching was reported in the 1960's and was based on oxides in a metal-ion-metal (MIM) framework with formation/dissolution (SET/RESET) of filament electrochemical
in nature~\cite{Dearnaley_1970}. Amongst ReRAM's, electrochemical metallization mechanism (ECM) or conductive bridging random access memory (CBRAM) which utilizes the
electrochemical dissolution of an active electrode material such as \Cu  or \Ag for SET/RESET mechanism has shown particular promise. In CBRAM, transition metals in 
their ionic state are converted to a conducting filament by applying a suitable electric field, which upon reversal, destroys the filament resulting in a resistive state.
Several possible candidates for solid electrolytes have been studied elsewhere~\cite{Perniola_2017,Chen_2017,Tsuruoka_2010,Watanabe_2010,
Anantram_2014,Sandhu_2015}. Amorphous tantalum pentoxide/amorphous tantala (a-\tantala ) has been investigated as a possible candidate for memory devices,
anti-reflection coatings and optical waveguides due to its high dielectric constant, high refractive index, chemical and thermal stability~\cite{Chen_1997,Martin_2016,KPPRL}.
\Cu-doped a-\tantala shows promising properties for CBRAM based memory devices.

Several experiments as well as calculations~\cite{Watanabe, Watanabe1,Masakazu_2010,Masakazu_2015} have been carried out to understand conduction mechanisms in \tantala materials. 
In these studies ~\cite{Watanabe,Masakazu_2010,Masakazu_2015}, \Ag, \Cu, $Pt$\hspace{0.09cm} metals were used as electrodes while in one \textit{Xiao et. al.}~\cite{Watanabe1} used \Cu  nanowires of 
different diameters inserted into the low density \tantala host to study transport and electronic properties of \tantala as an electrolyte. These papers indicate that 
metal filaments are responsible for conduction. It has been reported that conduction paths in 
different electrolytes differ qualitatively. Metals such as \Cu  form clusters, leading to a conducting filament in oxides, while no such clustering is observed in
chalcogenide based electrolytes~\cite{Kiran_PSS}. Since a complete investigation of \Cu -doped \tantala has not yet been reported, we provide here a thorough investigation amid growing research to test its candidacy as a possible electrolyte for CBRAM technologies. 

In this paper, we investigate the structural,
electronic and lattice dynamics of amorphous \Cu -doped a-\tantala. We provide insights into structural properties and coordination statistics, electronic and vibrational properties,
and visualize conduction/current paths by computing the space-projected conductivity (SPC)~\cite{Kiran_PSS}. We elucidate the atomistic mechanisms of phase segregation and track the emergence of \Cu clusters as the melt cools. The rest of paper is organized  as follows. In section II, we discuss the computational methodology used to generate our models.
This is followed by  validation of the generated models with particular attention to the structural, electronic, vibrational, and thermal 
properties in section III.  In Section IV we present the conclusions of our work and future research directions.


\begin{figure*}[!ht]
\begin{minipage}[b]{.475\textwidth}
  \centering
  \caption*{\Large\textbf{(a)}}
  \includegraphics[width=0.90\textwidth]{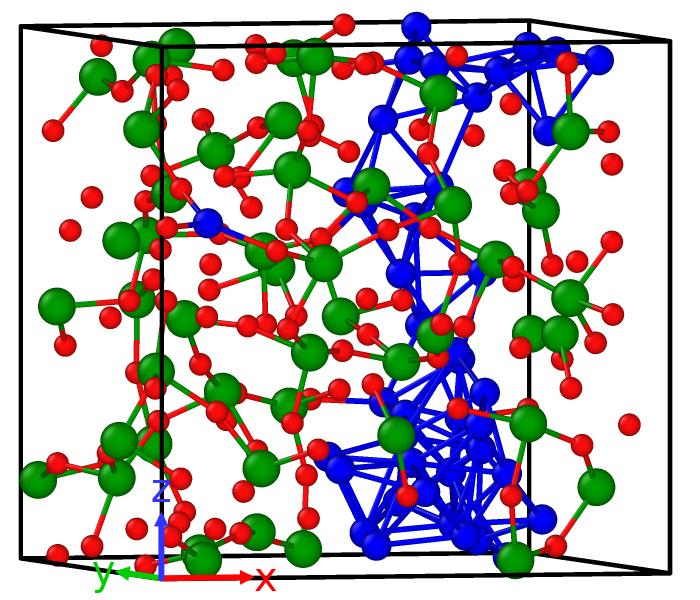}
\end{minipage}\hspace{0.2cm}
\begin{minipage}[b]{.475\textwidth}
\centering
\caption*{\Large\textbf{(b)}}
  \includegraphics[width=0.90\textwidth]{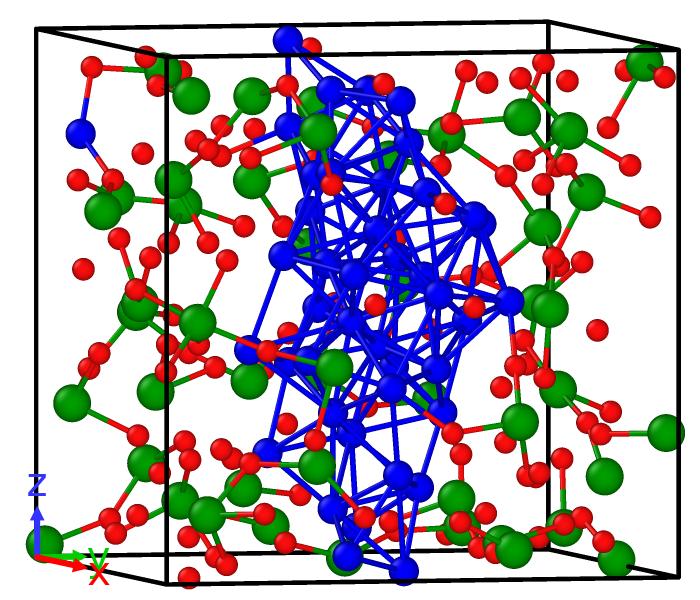}
\end{minipage}
\vspace{0.4cm}
\caption{\textbf{(a)} The structure of Model I \textbf{(a)} and Model II \textbf{(b)} consists of tantalum atoms bonded to 5, 6 and 7 oxygen atoms to form a mixture of edge-sharing, corner-sharing and face-sharing polyhedra and a connected subnetwork of \Cu atoms. The \Cu network grows in the interstitial space between $Ta$-$O$ polyhedra.  \Cu , \Ta and \O atoms are shown in blue, green and red, respectively and the same "color nomenclature" will be used throughout the paper.}
\label{fig:fig1}
\end{figure*}

\section{\label{sec:level2} Methodology and Models}

We prepare two $210$-atom models of a-$(Ta_2O_5)_{0.80}Cu_{0.20}$ cooled at different rates by utilizing melt-quenching within the \textit{ab initio} molecular dynamics (AIMD) method. 
The initial density for both of the models was chosen to be that of amorphous tantala ($\rho= 7.79 $ $gm/cc$), consistent with experimental~\cite{Bassiri1,Weber_2018} 
and theoretical studies~\cite{Watanabe1}. We have performed molecular dynamics simulations using \textit{ab initio} plane wave code VASP~\cite{vasp,vasp1,vasp2} with projector-augmented 
wave (PAW) method~\cite{PAW} and employed the \textit{Perdew-Burke-Ernzerhof} (PBE)~\cite{PBE1996} exchange-correlation functional.  Due to the size of the unit cell, only the $\Gamma$-point is used for Brillouin zone (BZ) integrations. A plane-wave cut off of 500 eV, a time step of 3.0 fs and Nose' thermostat was used to control the temperature.\\

\subsection{Model I}
We fabricated a starting model of 48 \Ta, 120 \O and 42 \Cu atoms with random initial positions (with no atoms closer than 2 $\mathring{A}$) in a cubic box of side 14.14 $\mathring{A}$. This model was then taken through a \textit{melt quench} (MQ) \citep{DAD_EPJB} cycle. Firstly, the system was heated well above melting point to form a liquid at 6000 K and then equilibrated at 6000 K, cooled to 3000K in 18 ps, equilibrated at 3000K for 9ps and further cooled to 300K in 15 ps summing up for a total simulation time of 57 ps. The cell volume was relaxed to obtain zero pressure models. This zero-pressure relaxation produced a volume rise of $2.09\%$ yielding an optimized density of $7.63$ $gm/cc$.

\subsection{Model II}
Another melt and quench (MQ) model, with slower cooling rate around the melting point of \tantala , was made. This model started with the melt of Model I cooled to 3000K and  was further cooled to 300K in 24ps without any intermediate equilibration. The total simulation time was 60 ps.

After dynamical arrest, conjugate-gradient relaxation was applied until the magnitude of the force on each atom was reduced to less than $0.01$ $eV/\mathring{A}$. Zero-pressure relaxation increased the volume by $2.39\%$ and the density was optimized to $7.61$ $gm/cc$. A third model, cooled faster than the models discussed here, has been described in the Supplementary Material, to provide some insight into the influence of the quench rate on the network topology.

For simplicity and consistency, we follow the same "color nomenclature" for the atomic species: \Ta , \O , and \Cu atoms are colored green, red, and blue respectively.
\begin{figure*}[!ht]
\begin{minipage}[b]{.7\textwidth}
\centering
  \includegraphics[width=1.02\textwidth]{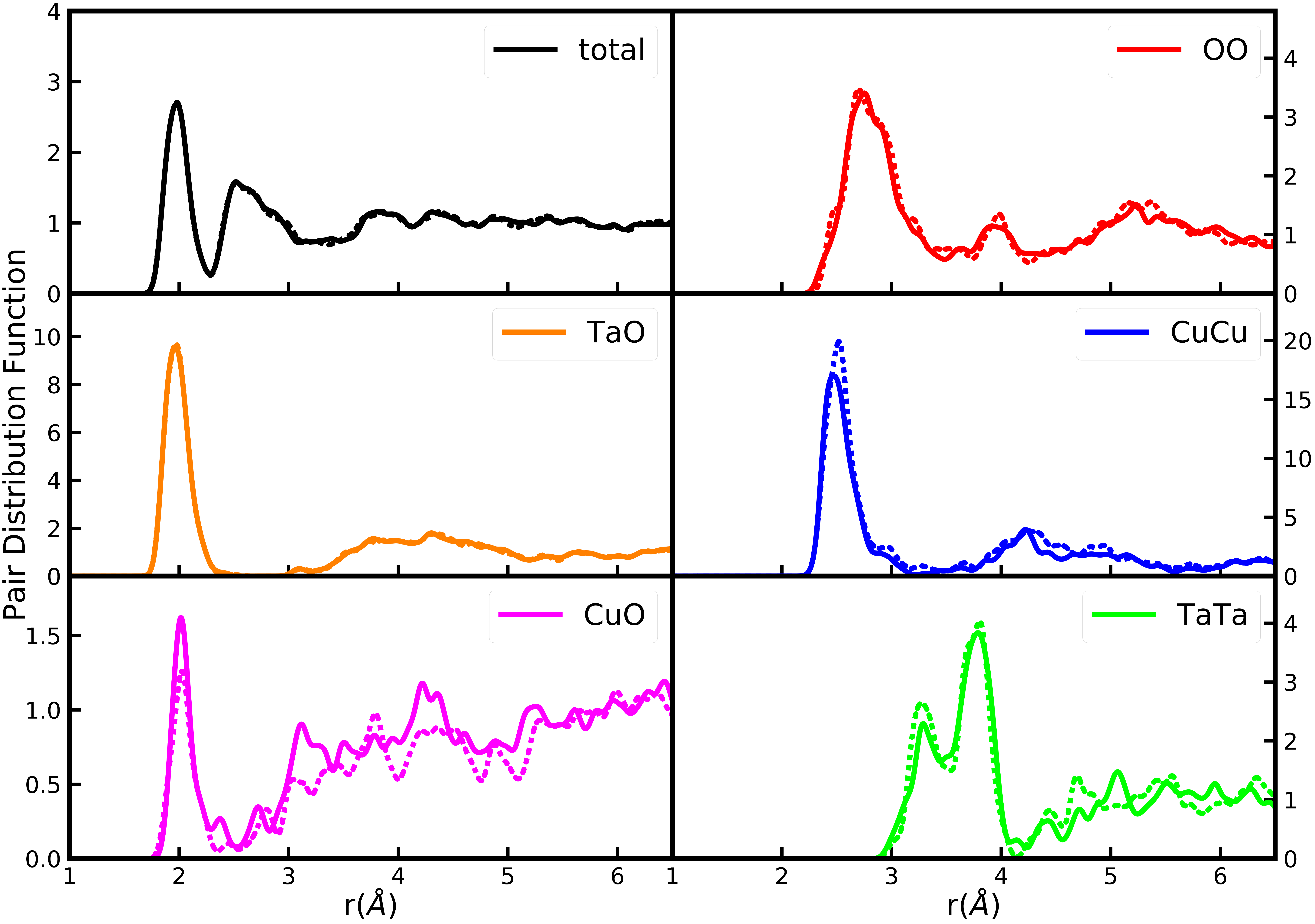}
\end{minipage}
\caption{Pair distribution function for Model I  (\textbf{solid line}) and Model II (\textbf{dotted line}). The total pair distribution function \textbf{(g(r))} is almost the same for both the models. There is, however, a slight change in the \Cu-\Cu and \Cu-$O$ partial correlations.}
\label{fig:fig2}
\end{figure*}

\begin{table}[htp]
\setlength{\arrayrulewidth}{0.60mm}
\centering
\begin{tabular}{|c| c |c |c |c |c |}
\hline
&Atom & $n$ & $n(Ta)$ & $n(O)$ & $n(Cu)$ \\ \hline
\rule{0pt}{4mm}
&$Ta$ & 7.96 & 1.75 & 5.52  & 0.69   \\
\rule{0pt}{4 mm}
$Model  I$&$O$ & 2.44 & 2.21 & 0.00 & 0.23 \\
\rule{0pt}{4 mm} 
&$Cu$ & 7.17 & 0.79 & 0.67 & 5.71 \\
\hline
&$Ta$ & 8.00 & 1.92 & 5.60  & 0.48   \\
\rule{0pt}{4 mm}
$Model  II$&$O$ & 2.45 & 2.24 & 0.00 & 0.21 \\
\rule{0pt}{4 mm} 
&$Cu$ & 7.48 & 0.55 & 0.60 & 6.33 \\
\hline
\end{tabular}
\caption{Average coordination number $(n)$ and its distribution among different species. Coordination is counted only if the distance between the atoms is no more than the sum of their covalent radius within a tolerance of 0.1 $\mathring{A}$. Covalent radii for \Ta, \O and \Cu are taken as $1.70 $\AA, $0.73 $\AA, and $1.32 $\AA, respectively}
\end{table}

\section{\label{sec: level4} RESULTS AND DISCUSSION}

\subsection{\label{sec:level4a} Structural Properties}

As shown in Fig.2, the radial distribution function for both models has a first peak at 1.95$\mathring{A}$, which arises from the dominant \Ta-\O bonds and corresponds to the \Ta-\O bond length. This peak is in agreement with experiments ~\cite{BadriShyam,Bassiri1} for pure \textit{a-$Ta_2O_5$} and previous DFT calculations ~\cite{Watanabe} for \Cu-doped tantala as well which suggests that the introduction of \Cu to the network does not significantly change the local environment around the $Ta$ atoms, i.e. the dominance of $Ta$-$O$ octahedra in the structure persists even after \Cu doping. This finding is also supported by low $Cu$-$O$ coordinations. Introduction of \Cu, however, steals some \O coordination from \Ta , as seen in Table I, and these under-coordinated \Ta atoms have a significant role in conduction which will be explained later. 

\begin{figure*}[!ht]
\begin{minipage}[b]{1.0\textwidth}
  \centering
  \includegraphics[width=0.90\textwidth]{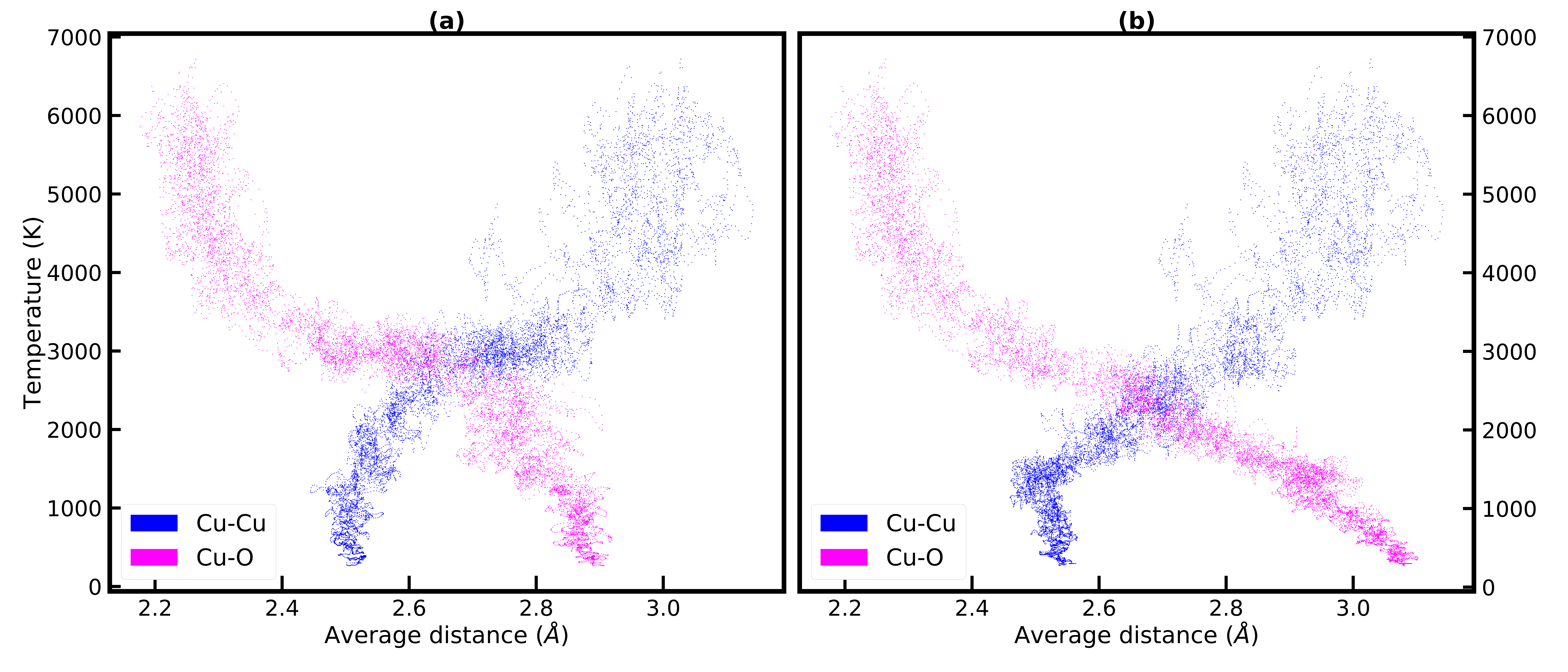}
\end{minipage}\hspace{0.2cm}
\caption{Average distance of three nearest \Cu to \Cu atoms (blue dots) and \O to \Cu atoms(purple dots), averaged over all \Cu atoms, for Model I \textbf{(a)} and Model II \textbf{(b)}.}
\label{fig:fig3}
\end{figure*}

The $Ta$-$Ta$ correlations also remain unaffected by \Cu-doping but a strong peak, at 2.48$\mathring{A}$, in the \Cu-\Cu correlations suggests the formation of \Cu clusters in the system, as seen in Fig.\ref{fig:fig1}, which can be attributed to the ionicity of the $Ta$ -$O$ bonds ~\cite{Kim} that drives the \Cu-atoms to cluster. Previously, clustering of \Cu in ionic host (a-$Al_2O_3$) has been reported ~\cite{Kiran_PSS}. This strong \Cu-\Cu correlation suggests that \Cu-atoms preferentially bond with themselves, consistent with the coordination statistics. In contrast, zirconia-doped tanatala shows no $Zr$ clustering and, the metal atoms distribute themselves homogenously with no metal-metal pair closer than 2.9$\mathring{A}$ \cite{KPPRL}. It is quite interesting that our calculations ``naturally" produce connected \Cu "wires" that are extended in space (considering the periodic boundary conditions), not by modeler's ``installation", but as a direct and unbiased consequence of the melt-quench simulations themselves.

\begin{figure*}[!ht]
\begin{minipage}[b]{.475\textwidth}
  \centering
  \caption*{\Large\textbf{(a)}}
  \includegraphics[width=0.90\textwidth]{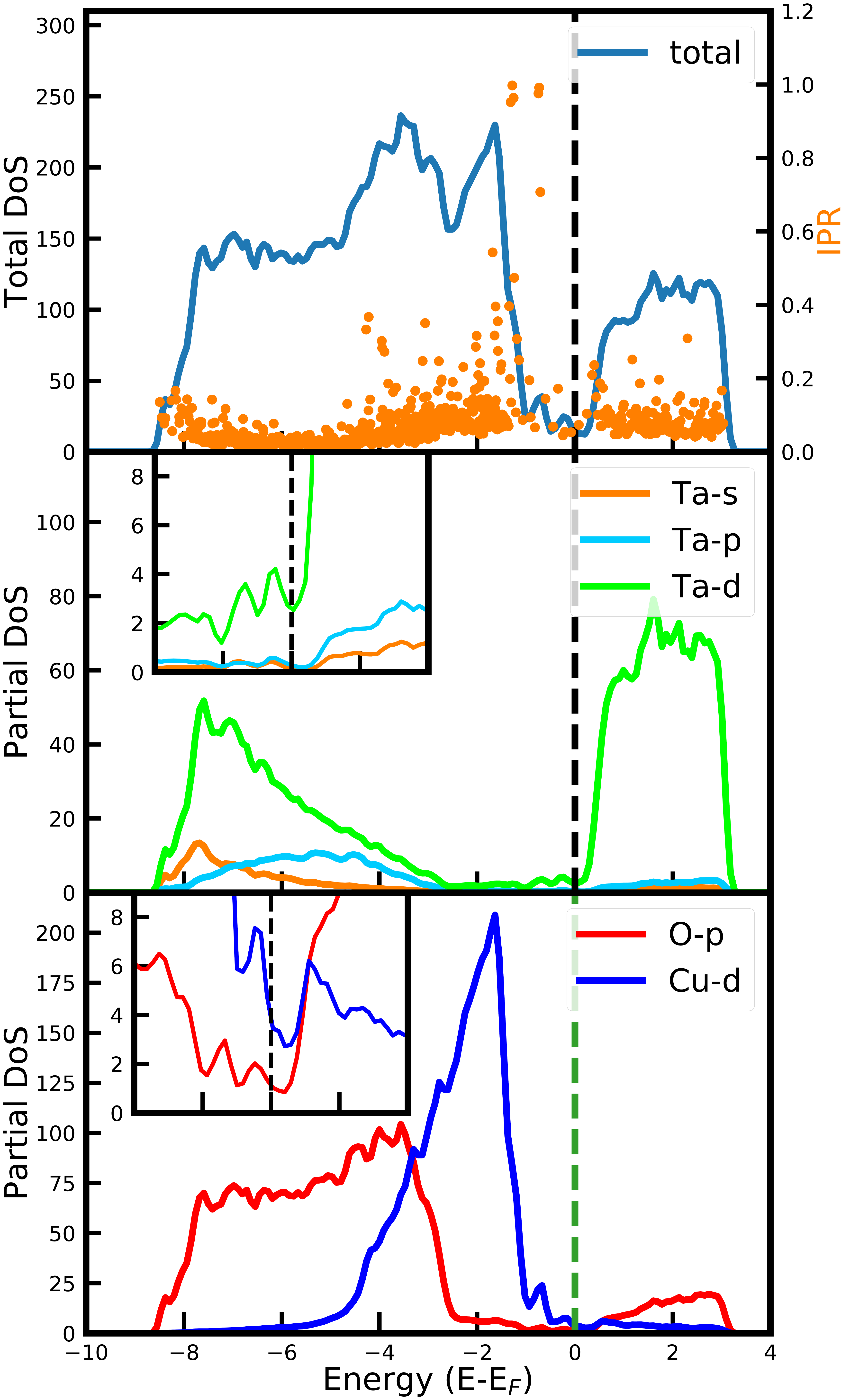}
\end{minipage}\hspace{0.2cm}
\begin{minipage}[b]{.475\textwidth}
\centering
\caption*{\Large\textbf{(b)}}
  \includegraphics[width=0.90\textwidth]{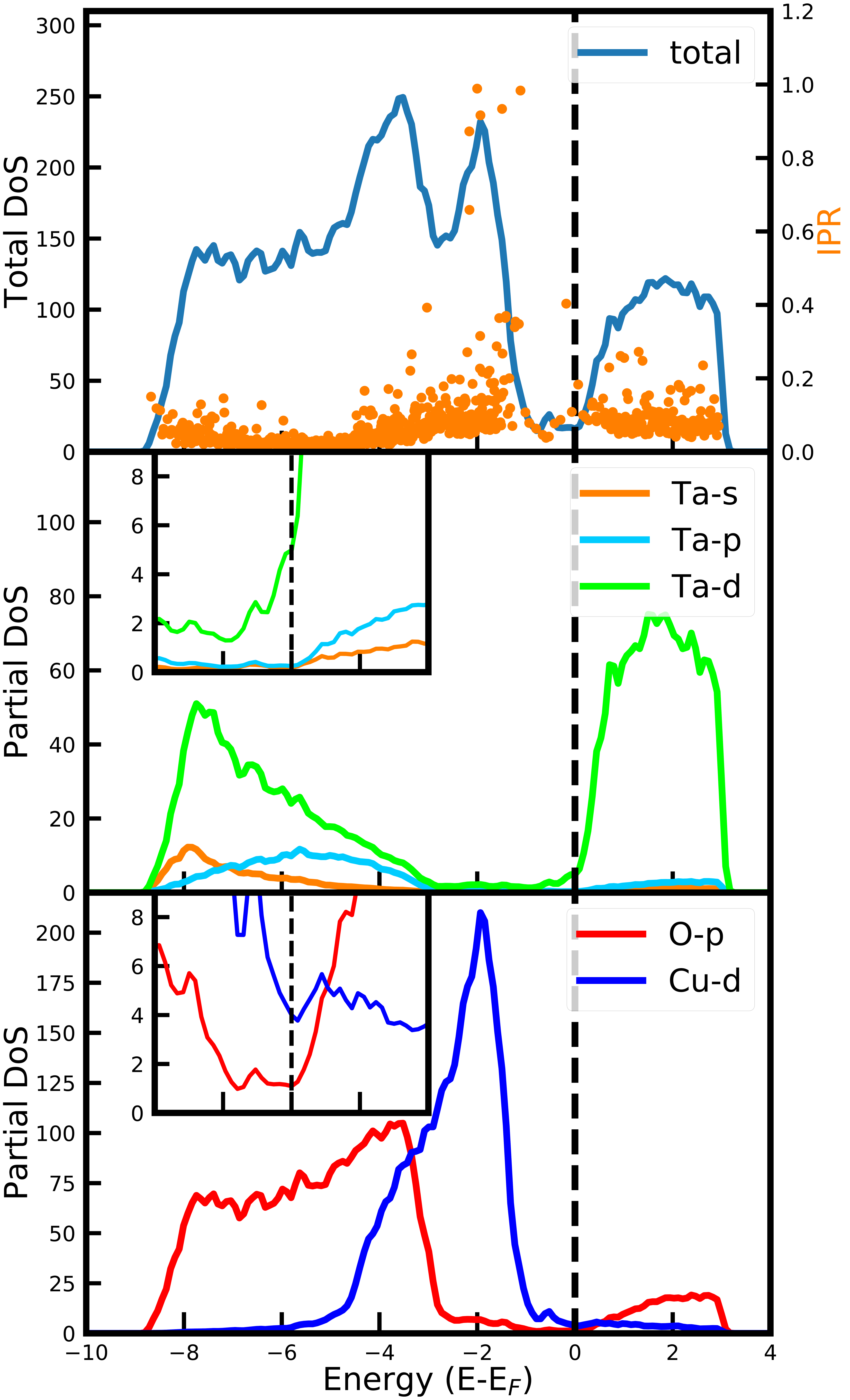}
\end{minipage}
\caption{Electronic density of (Kohn-Sham) states, Inverse Participation Ratio and projection onto atomic species for Model I \textbf{(a)} and Model II \textbf{(b)} with black vertical drop lines indicating Fermi level. The insets show a magnified version of the DoS contributions from each species near the Fermi Level.}
\label{fig:fig4}
\end{figure*}

Two peaks are worth mentioning in the $Ta$ -$Ta$ correlation: the first at around 3.3$\mathring{A}$, and the second around 3.8$\mathring{A}$, firstly because they provide an idea of the how the \Ta-\O octahedra are connected, and secondly because they are implicated in mechanical loss for Laser Interferometer Gravity Wave Observatory application~\cite{KPPRL}. The former comes from the joint contribution of face and edge-shared octahedral connection while the latter derives from the corner-shared connection of the octahedra. 

The coordination statistics of the models in Table I largely serve to validate the findings of the RDF and the correlations between different species. It also suggests that, as we lower the cooling rate, the \Cu-\Cu coordination increases while the \Cu-\O coordination decreases. This implies slower cooling rates produce better \Cu clusters with \O being pushed away from them. 

To compare the \Cu-\Cu and the \Cu-\O environment and its evolution during the \textit{melt and quench} process in our models, we calculate, at each step, the following quantities:
\begin{equation}
	d_{CuCu} = \frac{1}{3N_{Cu}} \sum_{i=1}^{N_{Cu}} \sum_{j=1}^{3}|{\vec{R}_{Cu,i}-\vec{R}_{Cu,j}}|
\end{equation}

\begin{equation}
	d_{CuO} = \frac{1}{3N_{Cu}} \sum_{i=1}^{N_{Cu}} \sum_{j=1}^{3}|{\vec{R}_{Cu,i}-\vec{R}_{O,j}}|
\end{equation}
The sum over \textit{j} runs over three nearest \Cu in the first equation and over three nearest \O in the second.  The scatter plots in Fig. 3 show how these distances change as we cool the melt. This  plot clearly shows that as we reduce the cooling rates, the \O and \Cu atoms move apart.  Furthermore, the details of the change in the bonding environments of the atoms and the phase segregation of \Cu atoms in the network during the \textit{melt and quench} process has been discussed with animations in the Supplementary Material. There, we provide a visualization of the network formation process, and observe the exclusion of \Cu as the host \tantala network, rendering the \Cu becomes {\it atomus non grata} in or near the \tantala regions. The main ``takeaway" from Fig. 3 is that the more extended cooling produces a more compact \Cu cluster for Model II (hence the extended right ``leg" on the right side of the ``ballerina plot" of Fig. 3b compared to Fig 3a). This suggests that slower cooling rates create \Cu clusters that are as compact as possible, and minimize \Cu cluster surface area exposed to the \tantala host. Of course this hints at a propensity to form crudely spherical clusters, though our simulations are too small to prove this assertion.

\begin{figure*}[!ht]
\begin{minipage}[b]{.245\textwidth}
  \centering
  \caption*{\Large\textbf{(a)}}
  \includegraphics[width=1.09\textwidth]{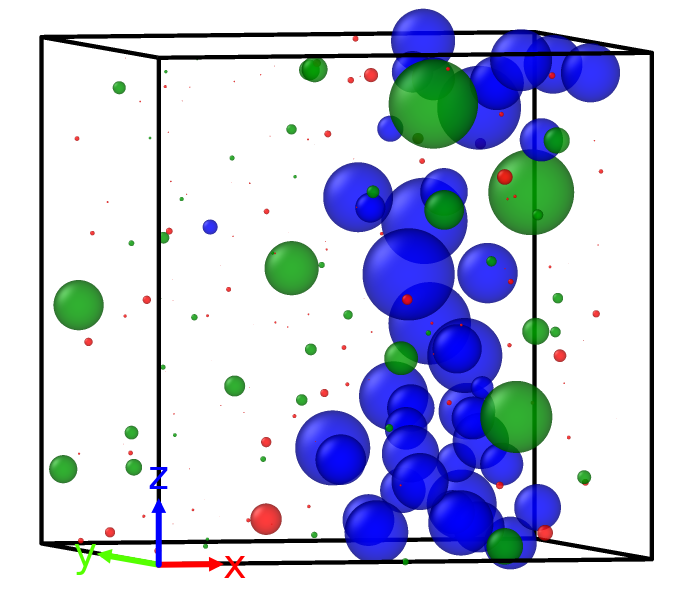}
\end{minipage}
\begin{minipage}[b]{.245\textwidth}
  \centering
  \caption*{\Large\textbf{(b)}}
  \includegraphics[width=1.09\textwidth]{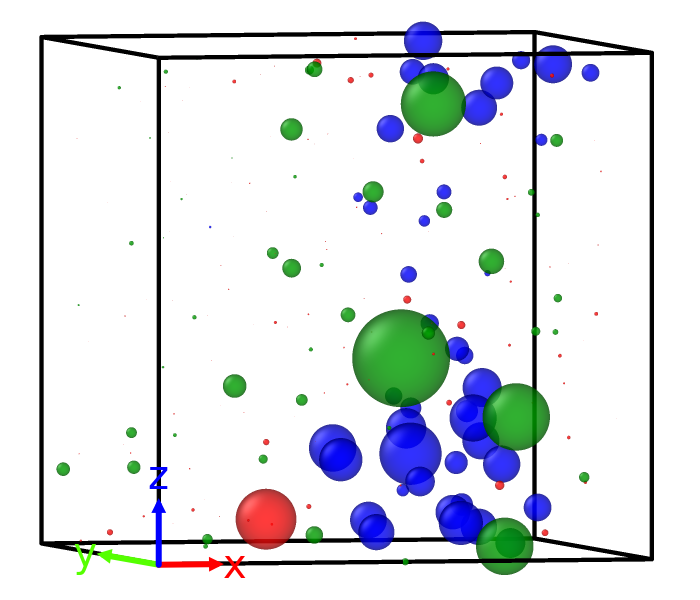}
\end{minipage}
\vspace{0.4cm}
\begin{minipage}[b]{.245\textwidth}
  \centering
  \caption*{\Large\textbf{(c)}}
  \includegraphics[width=1.09\textwidth]{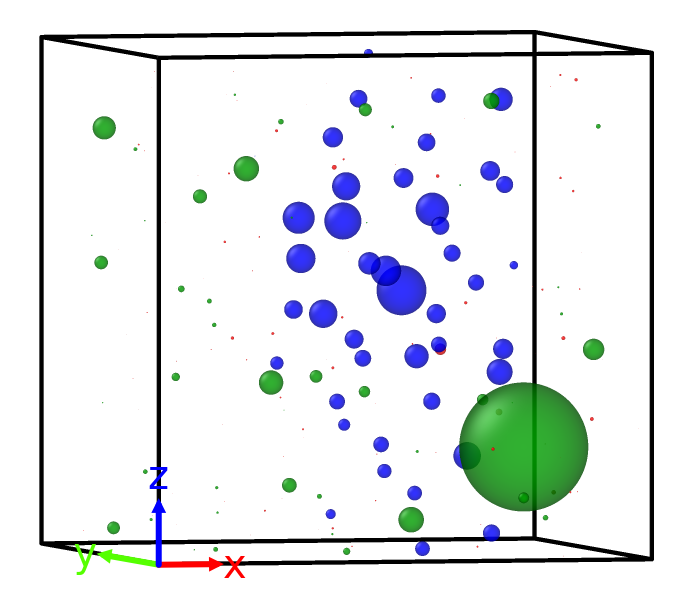}
\end{minipage}
\begin{minipage}[b]{.245\textwidth}
  \centering
  \caption*{\Large\textbf{(d)}}
  \includegraphics[width=1.09\textwidth]{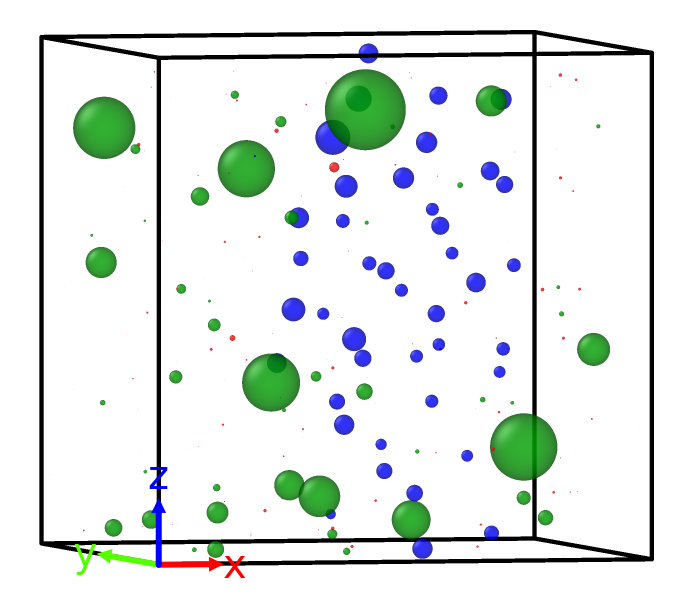}
\end{minipage}
\label{fig:fig5}
\caption{Electronic DoS averaged over three bands below (\textbf{(a)} for Model I and \textbf{(c)} for Model II) and above (\textbf{(b)} for Model I and \textbf{(d)} for Model II) the Fermi level. The size of the atoms is proportional to their contribution to the total DoS. Colors as in Fig. 1.}
\end{figure*}

\begin{figure*}[!ht]
\begin{minipage}[b]{.475\textwidth}
  \centering
  \caption*{\Large\textbf{(a)}}
  \includegraphics[width=0.90\textwidth]{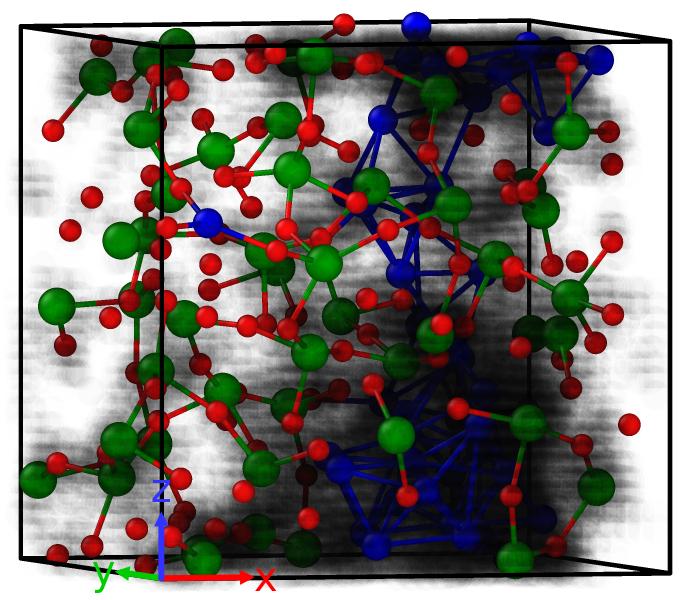}
\end{minipage}\hspace{0.2cm}
\begin{minipage}[b]{.475\textwidth}
\centering
\caption*{\Large\textbf{(b)}}
  \includegraphics[width=0.90\textwidth]{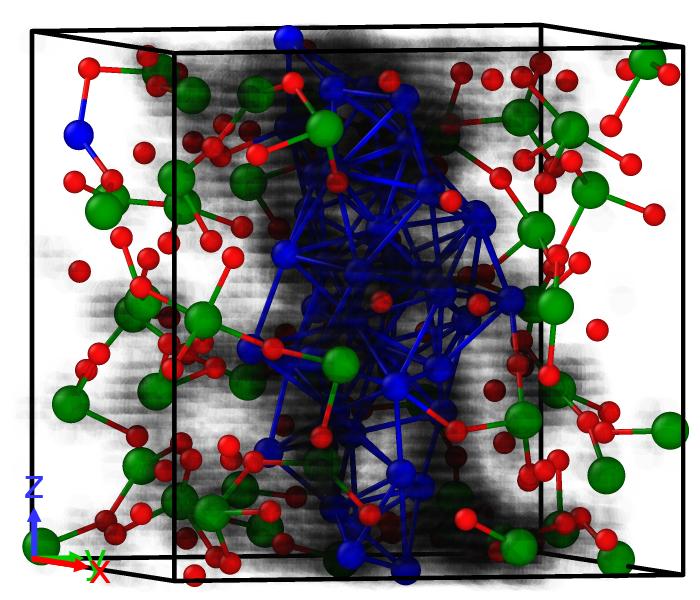}
\end{minipage}
\vspace{0.4cm}
\caption{Space Projected Conductivity scalar field for Model I\textbf{(a)} and Model II\textbf{(b)} shown in grayscale. Those parts of the network containing the interlinked \Cu-atoms are clearly more conducting in both models. The electrical conductivity of Model II is about 5 times that of Model I, because of the more robust \Cu filament of Model II.}
\label{fig:fig6}
\end{figure*}

\subsection{\label{sec:level4b} Electronic Properties}

To understand the electronic structure of the models, we examine the total density of states (DoS), partial DoS and  inverse participation ratio (IPR). These calculations not only help us check the validity of the model, but can also be used for \textit{a priori} information to model amorphous materials~\cite{Kiran1,Los1}. The plot of the DoS in Fig. 4 reveals that both models have states around the Fermi level with extended Kohn-Sham orbitals indicating conducting behavior. Since the host (\tantala) is an insulator with a band gap of $4.22 \textit{eV}$ ~\cite{WU1994}, we see that the introduction of \Cu to the network closes the gap by inducing impurity bands spread throughout the {\it entire} host (\tantala) gap. This is corroborated by the fact that the states near the Fermi level arise from the \Cu -3d orbitals hybridized mainly with \Ta and small contributions from \O orbitals, as seen in the partial DoS plots in Fig.4. 

Our calculations show that the states near the Fermi level arise mostly due to \Cu and \Ta and a small contribution from \O. The occurrence of \Cu clusters in the \tantala host suggest that the \Cu-clusters and \Ta atoms near them form the conduction-active parts. In order to study the details of the states near the Fermi level, we plot the species-projected DoS averaged over three states above and below the Fermi level in Fig.5. These plots show that states near the Fermi level arise from both \Cu and \Ta atoms and is shared among a fairly large number of atoms.  

A further insight into the electronic properties is given by IPR defined as:
\begin{equation}
 \mathcal{I}(\psi_n)= \frac{\sum_i {|a_{n}^i|}^4 }{\big (\sum_i {|a_{n}^i|}^2\big)}
\end{equation}
with $a_n^i$ being the contribution to the eigenvector $\psi_n$ from the $i^{th}$ atomic orbital ($s$, $p$, and $d$) as obtained from VASP. In physical terms, IPR of electronic states is a measure of localization: localized state having high IPR value (ideally equal to $\mathcal{I}=1$) while a completely  extended state having a value of (1/\textit{N}), i.e. evenly distributed over \textit{N} atoms. 
Near the Fermi level, we observe low IPR indicating delocalized states and conducting behavior of the models.\\

\subsubsection{Space Projected Conductivity}
The density of states provides some hints about the species contributing near the Fermi-level; however, the conduction also depends upon the localization of their electronic states and momentum matrix elements between Kohn-Sham states near the Fermi level.  Recently, we have developed a spatial decomposition of the Kubo-Greenwood~\cite{Kubo,Greenwood} formula that provides information about conducting paths in real space. By introducing a discrete grid in space, we show that the quantity:

\begin{equation}
\zeta(\textbf{x}) = \left|\sum_{\textbf{x}} \Gamma (\textbf{x},\textbf{x}) \right |
\end{equation}

provides such information at the spatial grid point $\textbf{x}$ and for which:

\begin{equation}
\Gamma (\textbf{x},\textbf{x}^\prime) = \sum_{ij\alpha}g_{ij}\xi^{\alpha}_{ij}(\textbf{x})(\xi^{\alpha}_{ij}(\textbf{x}^\prime))^*  .
\end{equation}
Here, $g_{ij}$ is defined in \textit{Prasai et. al} ~\cite{Kiran_PSS} and 
$\xi^{\alpha}_{ij}(\textbf{x}) \equiv \psi^*_i(\textbf{x})p^\alpha\psi_j(\textbf{x})$ is a complex-valued function, 
$\psi_i(\textbf{x})$ is the $i^{th}$ Kohn-Sham eigenfunction and $p^\alpha = \frac{\hbar}{i} \frac{\partial}{\partial x_\alpha} ,(\alpha = x,y,z)$. We have used this approach to describe transport in a solid electrolyte material~\citep{Kiran_PSS} and \Cu -doped \textit{a-alumina}~\cite{KashiPRM}. In a mixed (insulating/conducting) system like ours only a few eigenvectors of $\Gamma$ characterize essentially all conduction in the system.




The SPC for both models is visualized as a grayscale plot in Fig.6. The figure shows that connected \Cu atoms form primary sites of conduction as expected. However, some \Ta atoms, which are near the \Cu atoms also contribute significantly to the electronic conduction. A detailed analysis of the bonding environment of these \Ta atoms show that they are under-coordinated with oxygen,i.e. have less than (or equal to) five \O bonds, a result that in agreement with previous works on non-stoichiometric tantala ~\cite{Bondi}. A detailed discussion of the bonding environment and the coordination statistics of these \Ta atoms has been made in the Supplementary Material. Furthermore, slower cooling rates produces higher \Cu-\Cu coordination and better connectivity, thereby enhancing conductivity. There is a factor of about 5 higher conduction in Model II than Model I, presumably because of the small ``neck" interlinking \Cu in Model I.

\subsection{\label{sec:level5} Vibrational Properties}

\begin{figure*}[!ht]
\begin{minipage}[b]{.475\textwidth}
  \centering
  \caption*{\Large\textbf{(a)}}
  \includegraphics[width=1\textwidth,height=12cm]{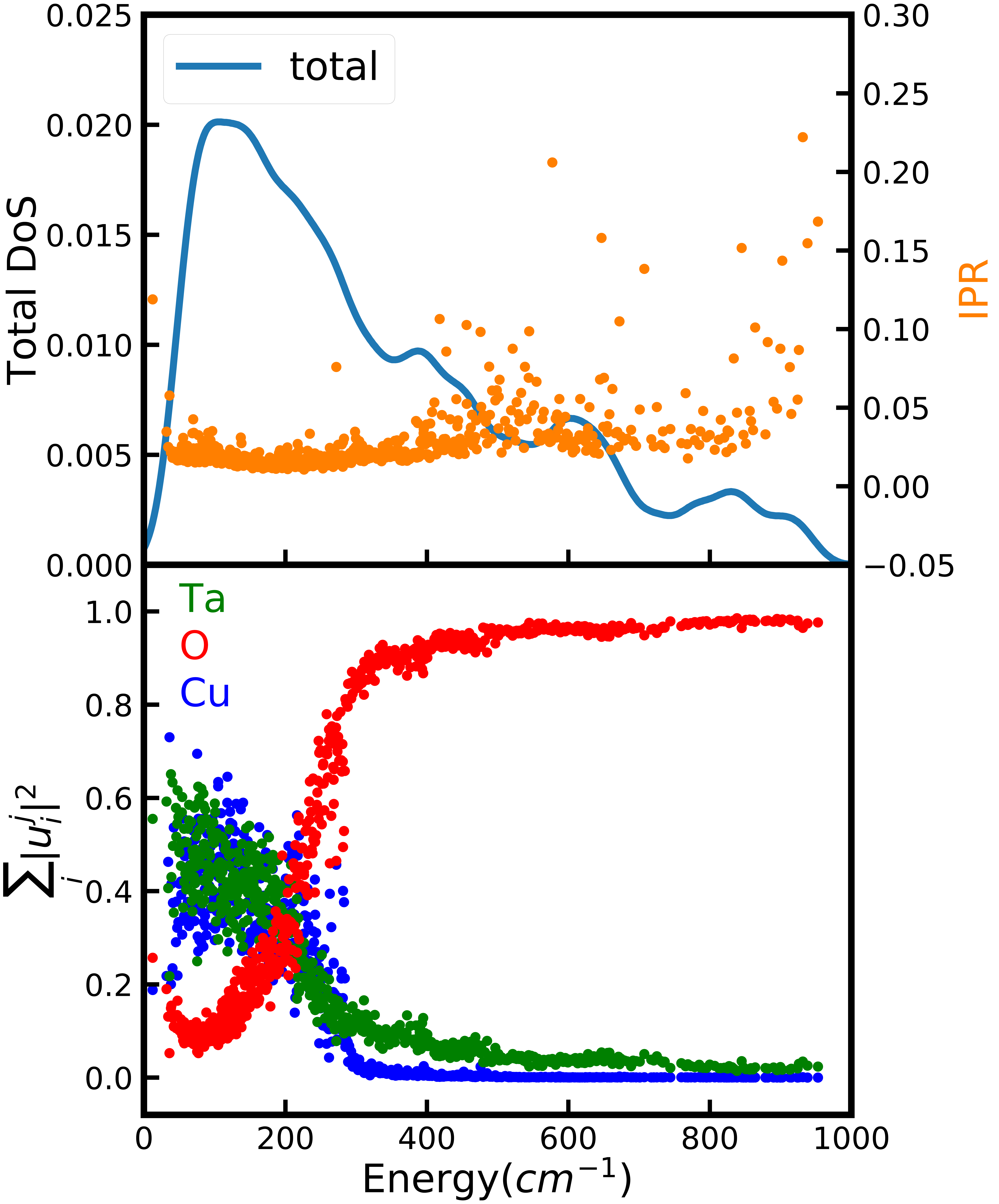}
\end{minipage}\hspace{0.2cm}
\begin{minipage}[b]{.475\textwidth}
\centering
\caption*{\Large\textbf{(b)}}
  \includegraphics[width=1\textwidth,height=12cm]{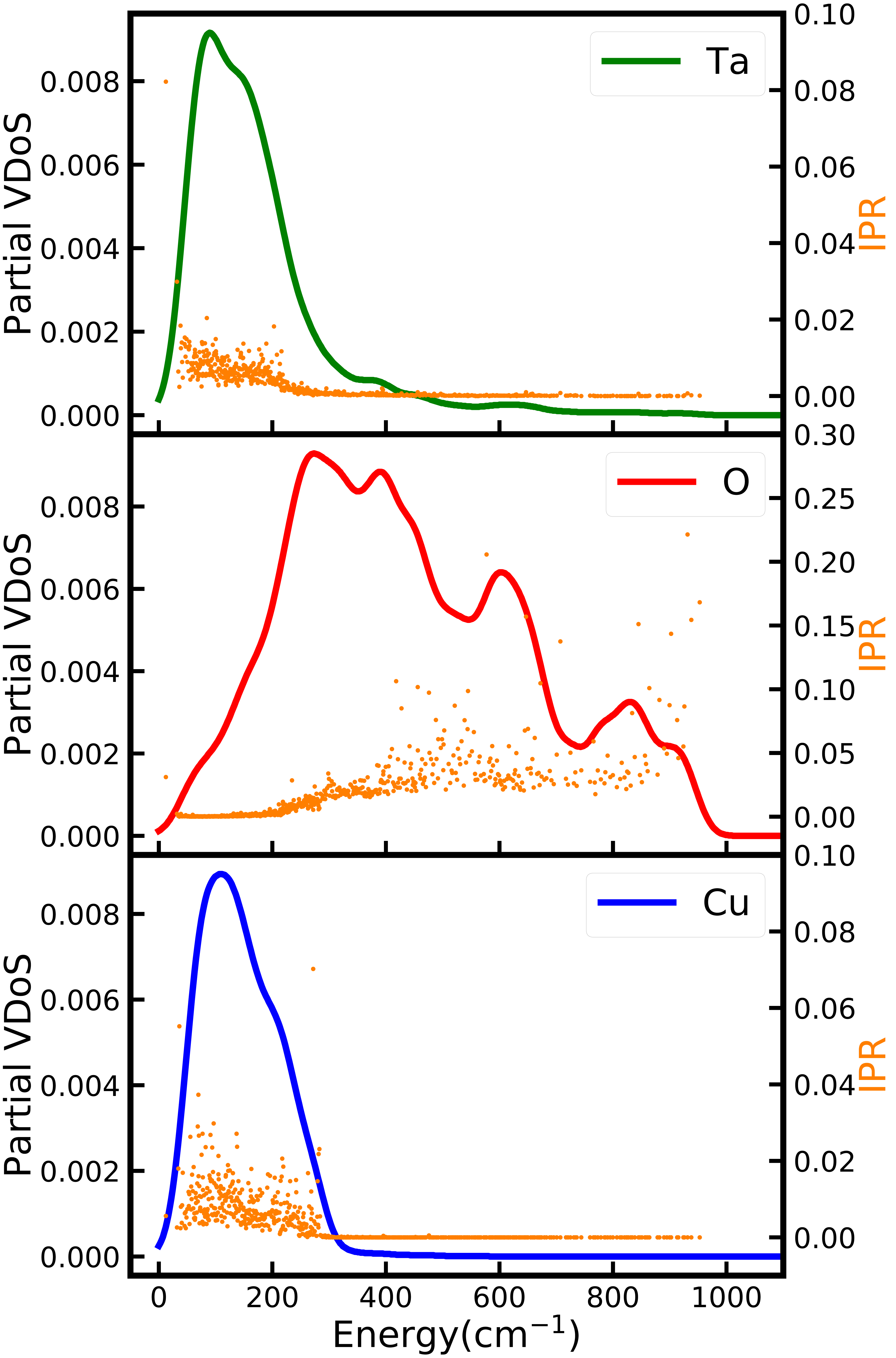}
\end{minipage}
\caption{ \textbf{(a, top panel)} Total vibrational density of states for the $Cu$ - doped \tantala (Model I) and the magnitude of normalized eigenvectors averaged over atomic species ($|u^j_i|^2$) in \textbf{(a, bottom panel)}. A transition is seen at frequency $\sim 270 cm^{-1}$ where the Oxygen atoms start dominating the vibrational spectrum.  The total vibrational localization (IPR) in \textbf{(a, top panel)} shows that phonons modes are mostly extended with few localized modes appearing at higher frequencies. The yellow circles show IPR that indicates localization of vibrational eigenmodes.  \textbf{(b)} We plot species-projected VDoS and VIPR of our Model I. We observe that oxygen dominates the higher frequency range and  the $Cu$  sub-network modes are mostly extended in nature. }
\label{fig:fig7}
\end{figure*}

\subsubsection{Vibrational Density of States}

The vibrational density of states (VDOS) provides key information about local bonding environments in amorphous solids and serves as a test to validate a model\cite{Lopinski1}.  Model I was well relaxed,  and the lattice vectors were simultaneously relaxed to attain zero pressure, which of course produces a slightly non-orthogonal supercell.  We displaced each atom in six directions($\pm x$,$\pm y$,$\pm z$) by ($\sim$ 0.015 $\mathring{A}$), and after each of these small displacements, forces were computed on all atoms, to obtain the force constant matrix, and dynamical matrix\cite{Bhattarai}. Classical normal modes were computed from the dynamical matrix by direct diagonalization. The VDOS is defined as:

\begin{equation}
g(\omega)=\frac{1}{3N} \sum_{i=1}^{3N} \delta(\omega-\omega_i)
\end{equation}

\noindent , with \textit{N} and $\omega_i$ representing the number of atoms and the eigenfrequencies of normal modes, respectively. 
To determine the elemental contribution to the VDOS, we computed species projected VDOS defined as ~\cite{Pasquarello}:
\begin{equation}
g_{\alpha}(\omega)=\frac{1}{3N} \sum_{i=1} ^{N_{\alpha}} \sum_{n} \lvert e_{i}^{n}\lvert^{2} \delta(\omega-\omega_n)
\label{Eq:VDOSPartial}
\end{equation}
\noindent , $ \lvert e_{i}^{n}\lvert^{2}$ are the eigenvectors of the normal modes and $N_{\alpha}$ is total number of atoms of $\alpha$ species.
These species-projected VDOS must satisfy the relation $g(\omega) = \sum_{\alpha} g_{\alpha}(\omega)$. 

As seen in Fig.7 (left panel), the VDOS is peaked at $\sim 105 cm^{-1} \approx 13 meV$, a peak arising due to the mixing of vibrational motion of  \Ta and \Cu atoms. Partial VDOS plot (right panel) shows that \Ta - and \Cu - vibrations are both peaked at $\sim 105 cm^{-1}$ while the 
\O atoms do not contribute to low frequency vibrations as significantly as the other species. However, at frequencies above $\sim 400 cm^{-1} $, VDOS contributions arise mainly from the \O -atoms, with no mixing, which can be ascribed to the low atomic mass of \O compared to \Cu - and \Ta - atoms. In the intermediate region ($200$ $cm^{-1}$ - $320$ $cm^{-1}$ ), vibrations arise from combined contributions of all atomic species. Animations of selected modes are provided in the Supplementary Material. Mode mixing and cross-talk between the phase-separated regions are features of these animations.

\subsubsection{Localization of vibrational modes}
While the VDOS is an observable that can be measured almost directly from inelastic neutron scattering experiments, the localization of these vibrations are not easily observable.
To study the localization of vibrational modes in the \Cu-doped \tantala, we calculate the vibrational IPR, the vibrational analogue of the electronic IPR, from the eigenvectors as shown in Equation.~\ref{Eq:VIPR}:

\begin{equation}
\mathcal{V}(\omega_n)=\frac{\sum_{i=1} ^N |\textbf{u}^i_n|^4 }{\big(\sum_{i=1} ^N |\textbf{u}^i_n|^2\big )}
\label{Eq:VIPR}
\end{equation}\vspace{0.2cm}
where ($\textbf{u}^i_n$) is displacement vector of $i^{th}$ atom at normal mode frequency $\omega_n$.

A small value of VIPR indicate evenly distributed vibration among the atoms while higher values imply only a few atoms contributing  at that particular eigenfrequency. We have plotted the total VIPR in Fig.7. Low values of VIPR below $\sim 300 \hspace{0.05cm}cm^{-1}$ suggest that the vibrational modes are completely delocalized/extended. Above $ 300 \hspace{0.05cm}cm^{-1}$, we observe higher VIPR. To provide visual insight to the spread of vibration over atoms and localization of some vibrational modes, suitable animations and explanations of some normal modes has been provioded in the Supplementary Material. 

To investigate the relation between the vibrational localization and atomic species, we evaluate contribution to VIPR from each atomic species, commonly called species-projected VIPR~\cite{Rajen_2019}. These projections sum up to the total VIPR, i.e. satisfy the relation:

\begin{equation}
\mathcal{V}(\omega_n)=\mathcal{V}_{Ta}(\omega_n)+ \mathcal{V}_{O}(\omega_n)+ \mathcal{V}_{Cu}(\omega_n)
\label{Eq:VIPRProjected}
\end{equation}\vspace{0.2cm}

\noindent and is shown on the right panel of Fig. 7. The species-projected VIPR calculations suggest that low frequency modes arise mainly from \Ta- and \Cu-atoms while the high frequency vibrations come mostly from the \O-atoms which can be attributed to the atomic masses of the species. Higher values of partial VIPR are seen at higher frequencies. Therefore, the high frequency modes are localized on a few \O atoms in the network while the low frequency modes are spread among larger number of \Ta and \Cu atoms. The quantity plotted in the bottom left panel Fig. 7 is the squared-magnitude of normalized eigenvectors summed across the atomic species for all the normal modes.  The scatter plot and the partial VDOS plots suggest that the \Ta- and \Cu-atoms participate almost equally in the low frequency vibrations. 

\section{\label{Comparison} Conclusions}
We describe the atomistic process of phase segregation of \Cu in a-\tantala. The  \Cu did not significantly alter the \Ta-\O bonding but instead phase separated, forming \Cu clusters. Models made with a slower cooling rate revealed  significantly better (denser) clustering than the one with faster cooling rate. These clusters, along with the neighboring under-coordinated \Ta atoms, form a conducting path in the network which is in agreement with previous literature, though presented in novel way in this paper, and not relying only on the Kohn-Sham states near the Fermi level, but also the momentum matrix elements, a legacy of the current-current correlation functions of Kubo. All this lends significant insight into an important CBRAM material. 

It is interesting to speculate on what would happen in larger models and different cooling rates. We might expect to see \Cu blobs in the network, possibly spatially separated but potentially interconnected by some other conducting fabric, perhaps Cu nanowires (of essential interest of course for CBRAM applications). While direct simulations like this one is computationally impossible for so large a system, it provides potentially useful {\it a priori} information for modeling employing simpler interatomic interactions. Electronic DoS calculations show that \Cu-doping closes the gap in the DoS of pure a-\tantala  with extended Kohn-Sham orbitals around the Fermi level. Vibrational modes at low frequencies are shared among many \Ta and \Cu atoms while those at high frequency are quite localized and come only from \O atoms. 

\section{\label{Comparison} Acknowledgements} 

We thank the National Science Foundation for support under grant DMR-1507670. We thank Dr. Kiran Prasai of Stanford University for helpful discussions. 

\bibliography{sample}

\end{document}